# Intricacies of $CO_2$-Basalt Interactions, Reactive Flow and Carbon Mineralization: Bridging Numerical Forecasts to Empirical Realities

Mohammad Nooraiepour*, Mohammad Masoudi, Beyene Girma Haile, Helge Hellevang
*Environmental Geosciences, Department of Geosciences, University of Oslo, Oslo, Norway*
* *mohammad.nooraiepour@geo.uio.no*

**Abstract**
Subsurface fluid flow and solute transport are pivotal in addressing pressing energy, environmental, and societal challenges, such as geological $CO_2$ storage. Basaltic rocks have gained prominence as suitable geological substrates for injecting substantial $CO_2$ volumes and carbon mineralization, driven by their widespread occurrence, high concentrations of cation-rich silicate minerals, reported fast mineralization rate, and favorable characteristics such as porosity, permeability, and injectivity. The mineralization process within basaltic rocks is intricately linked, involving the dissolution of silicate minerals and the subsequent precipitation of carbonate minerals. Columnar flow and batch surface growth experiments revealed the spontaneous formation of a limited number of large crystals at various locations, rationalized by the overarching influence of probabilistic mineral nucleation. Experiments with $CO_2$-acidified brine versus freshwater prove to be more challenging regarding the sweet spots for heavy carbon mineralization due to clay formation on the surface, particularly smectites. Despite numerical predictions suggesting the formation of MgFeCa-carbonates in $CO_2$-basalt interactions at higher temperatures, our laboratory findings primarily indicated the growth of calcium carbonates. The experimental and numerical outcomes highlight the necessity of a probabilistic approach for accurately modeling reaction kinetics, crystal growth distribution, and the dynamic interplay between reactive flow, geochemical reactions, mineral carbonation, and geometry alteration.

## Introduction

Subsurface fluid flow and solute transport are pivotal in addressing pressing energy, environmental, and societal challenges, such as the geological storage of carbon dioxide ($CO_2$). Basaltic rocks have emerged as highly suitable geological substrates for injecting large volumes of $CO_2$ with emission reduction and carbon mineralization purposes. This preference is attributed to their widespread occurrence at Earth's surface, high concentrations of cation-rich silicate minerals, reported fast mineralization rate, and often favorable characteristics such as porosity, permeability, and injectivity. The mineralization process within basaltic rocks is intricately linked, involving the dissolution of silicate minerals and the subsequent precipitation of carbonate minerals. During this chemical interplay, silicates play a crucial role by contributing vital calcium (Ca), magnesium (Mg), and iron (Fe) ions essential for the precipitation of carbonate minerals, including Ca-, Mg-, and Fe-carbonates (Hellevang et al. 2017. 2019).

Understanding the consequences of mineral growth in porous media and the fate of subsurface flow and transport necessitates spatial and temporal knowledge of solid precipitation locations and amounts (Nooraiepour et al., 2022). The reactive transport models can only then provide precise and realistic predictions on the intricate interplay between transport mechanisms and reaction kinetics and, therefore, advection-diffusion-reaction (ADR) (Masoudi et al., 2024). However, accurately representing the dynamics of mineral growth in porous media is still challenging. There is a continued need for theoretical development to precisely predict the occurrences where ADR coupling occurs in the space and time domains.

## Materials and Methods

**Flow-through column reactor experiments.** A specially designed glass tube reactor, 40 cm in length and 14 mm in inner diameter, was crafted and filled with crushed calcium carbonates (initial 4 cm) and basaltic glass (remaining 36 cm). Calcium carbonates ensured the requisite solute concentration ($Ca^{2+}$) upon exposure to $CO_2$-acidified liquid. The basaltic glass, sourced from Stapafell (Reykjanes Peninsula, Iceland), featured a tholeiitic composition, characteristic of a fine-grained extrusive igneous rock with relatively high silica and low sodium content. Basaltic glass was chosen for its rapid reactivity, homogeneous chemical composition, and easy identification of secondary carbonate phases due to its dark color. To prepare the experimental fluid, 1 liter of seawater from the Oslo Fjord underwent initial filtration through a 0.45 µm Millipore© filter. Milli-Q water provided the necessary deionized water (DI-water). $CO_2$-charged seawater was then prepared using a scientific-grade $CO_2$ bottle within a pressurized fluid accumulator housed in a forced convection benchtop oven (Fazeli et al., 2020), maintaining a gaseous pressure of 4 MPa. During the flow experiments, a controlled-rate injection protocol ensured DI-water/brine flow through the bottom port, facilitated by a dual-piston ISCO pump (setup details in Nooraiepour et al., 2018a,b). A steady-state flow of $CO_2$-charged liquid, equivalent to 0.005 ml/min was achieved. Fluid sampling occurred at the column inlet at discrete intervals (3 days apart) and continuously at the outlet. The acidity was analyzed post-sampling using a benchtop pH meter. The experiments were conducted at 80°C and atmospheric $CO_2$ pressure over 30 days.

**Batch-type microfluidic experiment.** A series of static (batch-type) experiments were carried out, involving variations in reaction temperature (80, 100, and 150°C) and aqueous solution pH (ranging from 5.7 to 9.86) for 21 to 52 days. The experimental solutions were prepared by diluting a 1 Molar HCl solution with Milli-Q DI-water. Separately, solutions of $CaCl_2.6H_2O$ and $MgCl_2.6H_2O$ were prepared by their dissolution in diluted HCl solutions, leading to distinct solutions with specific chemistries for each experimental run. For batch reactions, 10 mL Teflon-lined vessels were utilized, enclosed within an aluminum frame to prevent deformation and leaks. The experiments were conducted in aqueous solutions that were both supersaturated and undersaturated with respect to magnesium and calcium carbonate (Hellevang et al, 2017). Each experiment was repeated twice to ensure repeatability and reproducibility.

**Solid surface and aqueous phase characterization.** Scanning electron microscopy (SEM), backscattered (BSE) and secondary electrons (SE), was used to investigate surface structure and mineral growth. Chemical analyses and element mapping were conducted using energy-dispersive x-ray spectroscopy (EDS). High-resolution x-ray microcomputed tomography (micro-CT) was employed to extract the pore space structure of basalt samples. The aqueous solutions underwent analysis utilizing an ion chromatography system (ICS-2000). Silica concentrations were determined using an AutoAnalyzer 3. The total iron content ($Fe^{2+}/Fe^{3+}$) in the reacted bulk solution was determined through colorimetric analyses using a UV-VIS Spectrophotometer.

**Geochemical phase equilibrium modeling.** We used the PHREEQC v3 for aqueous geochemical calculations to compute/predict the solute saturation state, experimental pH, $CO_2$ pressure, and reaction progress. The adopted standard state was the unit activity for pure minerals and $H_2O$ at any temperature and pressure. A hypothetical one-molal solution was referenced to infinite dilution for all other aqueous species.

**Results and Discussion**

Large igneous provinces comprise massive magmas, rich in iron and magnesium, encompass continental flood basalts, volcanic passive margins, and related intrusive rocks. Within this category, continental flood basalts (Fig. 1) present an attractive $CO_2$ storage option owing to the internal fractures formed during lava cooling, enhancing the porosity and permeability of the rock matrix. Figure 1 illustrates the three distinct facies of Icelandic basalts, their x-ray tomograms, and SEM micrographs representing different scales of porosity (vascularity) within these rocks. The uppermost section of a basalt flow commonly features vesicles (Fig. 1d-i), providing a conduit for $CO_2$ plume migration and a suitable substrate for the progress of geochemical reactions leading to dissolution-precipitation. The subfigures juxtapose a tentative span of pore space characteristics and solid substrate properties in basalt reservoirs. As Figure 1 illustrates basaltic rocks may cover a broad range of eruptive styles, lithologies, and textural features, translating to a broad range of porosity (storage capacity) and permeability (injectivity). The diverse range of eruptive styles observed in basaltic volcanic eruptions is significantly shaped by the dynamics, timing, and length scales of vesiculation, gas percolation or escape, and bubble collapse. Examining vesicle metrics in basaltic rocks provides crucial insights into their parental magmas' percolation threshold and their respective petrophysical characteristics controlling fluid flow and solute transport.

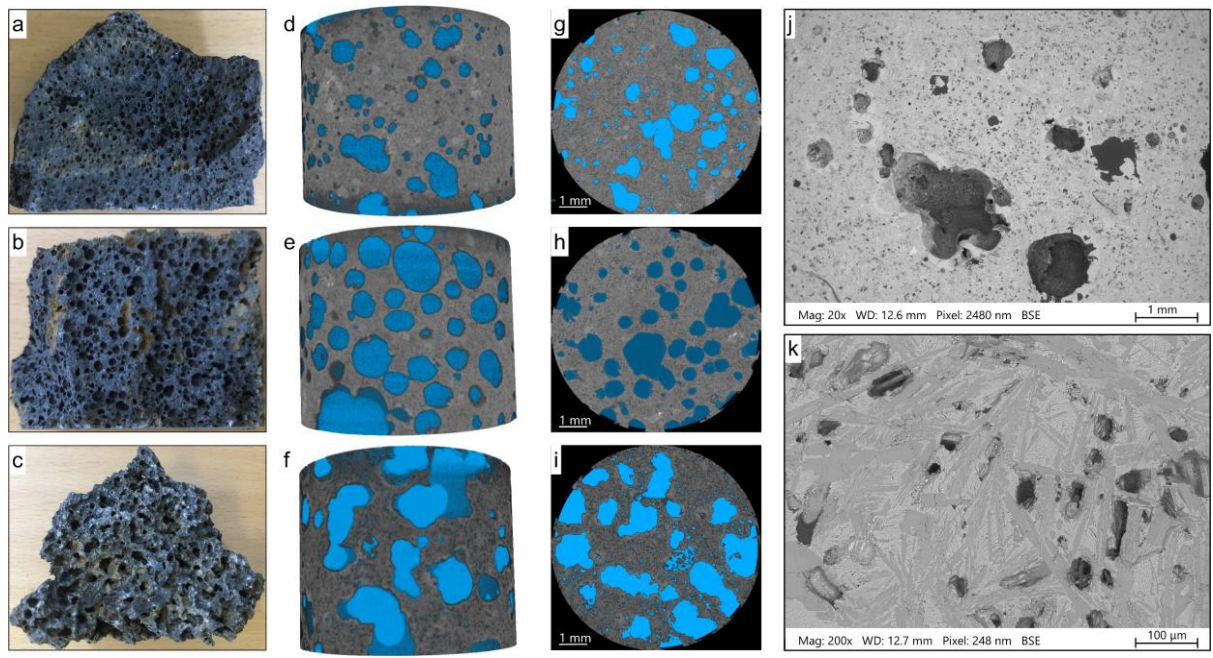

**Figure 1.** Sample characterization of three distinct sampled Icelandic basalt reservoir rocks (a-c) optical image, (d-i) segmented x-ray micro-computed tomography, (j-h) scanning electron microscopy with two different inspection scales depicting submicron porosities. The subplots visually convey the distinct properties of these three reservoir representative core samples, reflecting their facies and pore space geometries.

Figure 2a presents a schematic of the column reactor filled with calcite and basaltic glass during the flow-through experiments of $CO_2$-acidified freshwater/seawater. The precipitation of calcium carbonate minerals at two temporal instances (7 and 21 days) and two different sections of the column is illustrated. Early-stage non-crystalline shapes on the surface of secondary substrates (previously precipitated carbonates) are demonstrated using secondary electron imaging (Fig. 2d). Continuous monitoring of precipitation locations shows that the identified white patches appear random along the entire column length, specifically within the basalt section. Interestingly, larger accumulations occurred in the second half of the column, deviating from the numerical/modeling expectations based on a deterministic approach, which predicted accumulations at or close to the transition of calcite-basalt sections. The secondary precipitates formed as isolated pockets rather than dispersed aggregates throughout the column. These localized bodies of precipitates, their randomly spatial occurrences, and the fact that only a limited number but significantly larger crystal accumulations were formed collectively suggest that the overall precipitation reaction is predominantly controlled by nucleation rather than growth.

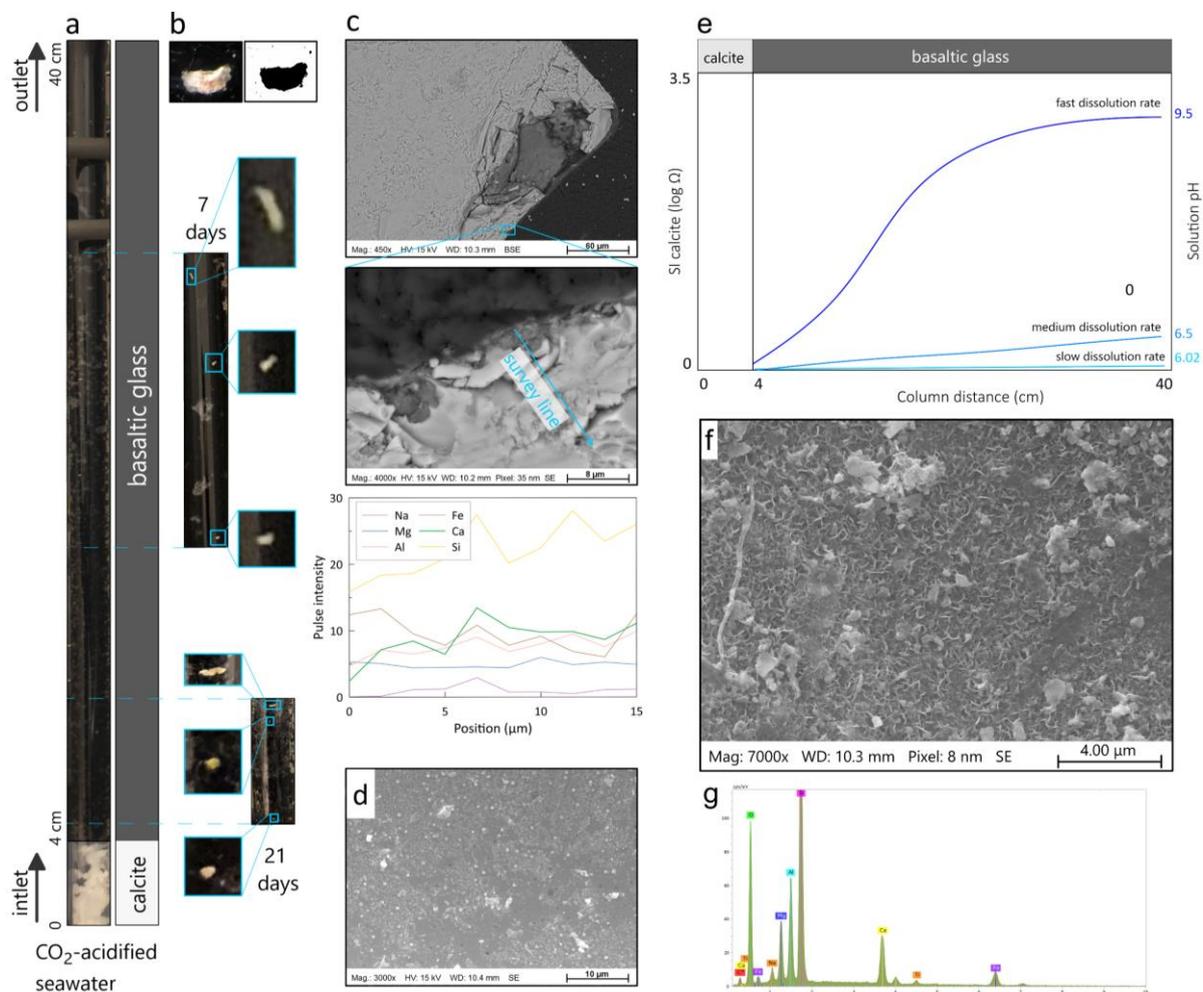

**Figure 2.** (a) Schematic representation of reactive flow-through tests. (b) Carbonate precipitation after 7 and 21 days at two different sections. (c) SEM and EDS analyses, depicting tracking of elemental concentrations on the substrate surface. (d) Visualization of nucleation events and their early-stage non-crystalline shapes on the substrate surface. (e) geochemical simulation of experimental column for 3 basalt dissolution rates along with the acidity of effluent experimental fluid. (f) surface growth of clays particularly smectites, (g) EDS spectrum of post-experiment specimen indicating increased surface Ca concentration.

Aqueous geochemical modeling indicated that upon entering the column, the injected fluid was buffered to a pH of approximately 6 in equilibrium with calcite before the subsequent dissolution of basaltic glass further buffered the solution. The measured pH of the outlet effluent solution exhibited a declining trend during experiments, from 7.25 to 6.58. This decline in pH over time could be attributed to either reduced basalt dissolution rates or increased precipitation of secondary carbonate phases with time. PHREEQC simulations complement this observation by illustrating that carbonate minerals remain stable, and the reacted aqueous solution becomes progressively more supersaturated (with respect to carbonates) as it advances through the column and reacts with the basalt substrate.

The dissolution of basalt led to the release of metal cations, potentially reacting with carbonate and forming solid carbonate minerals. Post-reaction SEM characterization of batch specimens revealed the formation of non-carbonate phases exhibiting a texture resembling clays, particularly smectites (Fig. 2f) and a carbonate phase (calcium-rich carbonate). In reactions occurring at lower pH levels (PHREEQC-estimated initial pHs of 5.89, 6.66, and 7.94), the presence of smectite overgrowths was evident, but there was no visible indication of carbonate crystals on the glass surface. At these pH levels, all the surfaces of the basaltic glass were covered with smectite. Conversely, experiments conducted with initial pH levels of 9.36 and 9.95 resulted in localized calcium carbonate crystals forming alongside smectite.

Regarding carbonate mineralization during $CO_2$-basalt interactions, geochemical modeling suggested the possibility of MgFeCa-carbonates formation, particularly at higher temperatures of 100 and 150°C. Contrary to these, our experimental observations solely revealed the growth of calcium carbonate polymorphs, namely calcite and aragonite. Ca-carbonates are not unexpected, as previous studies have demonstrated their presence during $CO_2$-basalt interactions across various temperature ranges. However, the absence of MgFeCa-carbonates, such as ankerite, siderite, and magnesite, remains unclear, but it is most likely closely linked to the concurrent formation of smectites, observed consistently across all experiments. Three potential explanations for the absence of these carbonates include: (i) the depletion of Mg and Fe from solution due to smectite growth; (ii) inhibition of nucleation due to the precipitation of a honeycomb smectite coating, creating an activation barrier; and (iii) prolonged growth rates caused by the large aqueous $Me^{2+}/CO_3^{2-}$ activity ratios. Kinetic models that do not account for the kinetic effects of nucleation and the activity ratios, such as many TST-derived equations, may inadequately predict the carbonatization potential in basalts. Smectites may also adversely impact fluid flow, where the growth of clay particles can rapidly clog pore throats and narrow fractures, leading to a fast decline in percolations.

In Figure 3 (top), the coordination number of pores in sandstone (conventional CCS reservoir) is compared to that in basalts (reservoir for unconventional CCS). The data is derived from the segmentation of x-ray tomography data. The 3D adjunct spaces in sandstones typically range from 4-6, whereas in basalts, they are predominantly limited to mostly 2, representing a one-to-one connection of vesicular bodies. This significant constraint in basalts markedly limits reactive transport, particularly where carbonate mineralization is prominent. We show percolation pathways in a high-permeability basalt by integrating micro-CT data and numerical simulations of flow percolations, when $CO_2$-induced carbonate precipitation and partial clogging occur (Fig. 3, bottom). The consequences depend on the location of precipitations, a probabilistic phenomenon. Small or large portions of flow channels can be affected, as illustrated in Fig. 3d (large precipitates with probabilistic precipitation location in two flow highways). Our pore-scale reactive Lattice Boltzmann methods reveal that even without prominent patches, numerous small precipitations can have more damaging consequences for flow and transport, contrary to the initial line of thought.

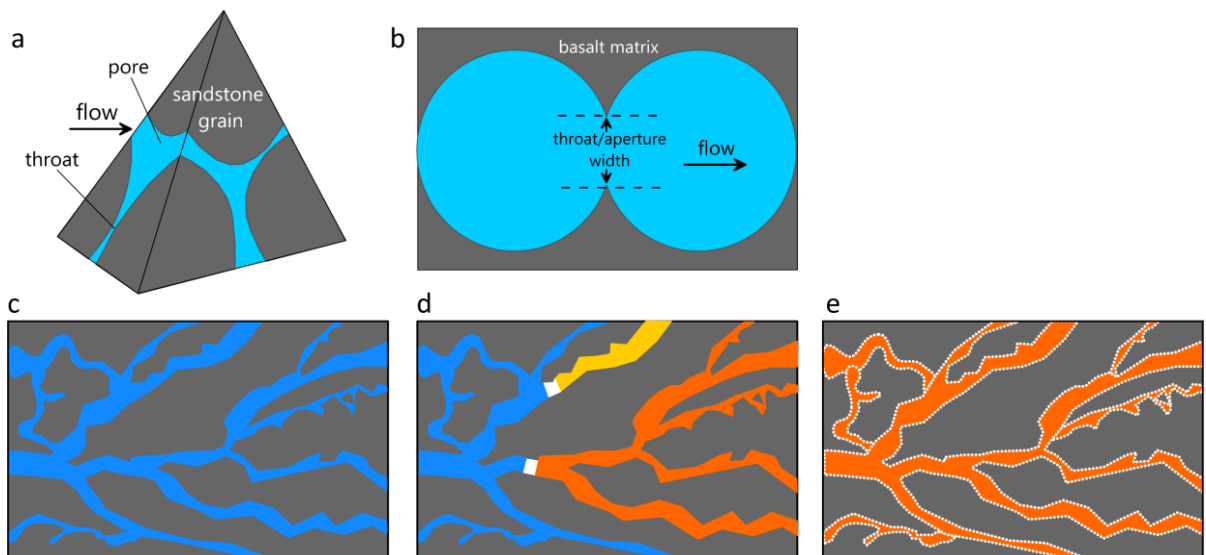

**Figure 3.** (top) Comparison of (a) sandstone and (b) basalt pore spaces, where a pore coordination number of 4-6 versus approx. 2 is expected, respectively. (bottom) (c) basalt percolation pathways (from micro-CT) along with (d-e) two clogging scenarios (big crystals and numerous small patches).

**Conclusions**

Columnar flow experiments revealed the spontaneous formation of a limited number of large crystals at various locations, rationalized by the overarching influence of probabilistic mineral nucleation. Experiments with $CO_2$-acidified brine versus freshwater prove to be more challenging regarding the sweet spots for heavy carbon mineralization due to clay formation on the surface, particularly smectites. Despite numerical predictions suggesting the formation of MgFeCa-carbonates in $CO_2$-basalt interactions at higher temperatures, our laboratory findings primarily indicated the growth of calcium carbonates. The experimental and numerical outcomes underscore the need for a new probabilistic approach to accurately model kinetics and crystal growth distribution, where dynamic ADR may steer geochemical reactions and subsurface reactive flow.

**Acknowledgements**

This research was supported by "Carbon mineralization in basaltic rocks: Understanding coupled mineral dissolution and precipitation in reactive subsurface environments" project, funded by the Norwegian Centennial Chair (NOCC).